\documentclass[runningheads]{llncs}
\usepackage[T1]{fontenc}
\usepackage{graphicx}

\usepackage{subcaption}
\usepackage{comment}
\usepackage{xspace}
\usepackage{todonotes}
\usepackage[inline]{enumitem}
\usepackage[linesnumbered,lined,boxed,commentsnumbered,noend]{algorithm2e}
\usepackage{stmaryrd}
\usepackage{amsfonts}
\usepackage{amsmath}
\usepackage{orcidlink}
\usepackage{tikz}
\usepackage[absolute,overlay]{textpos}
\definecolor{verifiergreen}{rgb}{0,.7,0}
\newcommand{\prdom}[1]{{\color{red}{#1}}}
\newcommand{\vedom}[1]{{\color{verifiergreen}{#1}}}
\newcommand{\pudom}[1]{{\color{blue}{#1}}}
\newcommand{\oncirc}[1]{\llbracket{#1}\rrbracket}
\newcommand{\prepro}[1]{\prdom{#1}}
\newcommand{\prever}[1]{\vedom{#1}}
\newcommand{\prepub}[1]{\pudom{#1}}
\newcommand{\postpro}[1]{\oncirc{\prdom{#1}}}
\newcommand{\postver}[1]{\oncirc{\vedom{#1}}}

\newcommand{\wire}[1]{\mathsf{wire}({#1})}
\newcommand{\assert}[1]{\mathsf{assert}({#1})}

\newcommand{\ZKSC}{ZK-SecreC\xspace}

\begin{document}

\title{Zero-Knowledge Proof-of-Location Protocols for Vehicle Subsidies and Taxation Compliance%
\thanks{\scriptsize This research has been funded by the Defense Advanced Research Projects Agency (DARPA) under contract HR0011-20-C-0083. The views, opinions, and/or findings expressed are those of the author(s) and should not be interpreted as representing the official views or policies of the Department of Defense or the U.S. Government.}}
\titlerunning{ZK-PoL for Vehicle Subsidies and Taxes}

\author{Dan Bogdanov\inst{1}\orcidlink{0000-0002-9296-9120} \and
Eduardo Brito\inst{1,2}\orcidlink{0009-0002-9996-6333} \and
Annika Jaakson\inst{1,2} \and
Peeter Laud\inst{1}\orcidlink{0000-0002-9030-8142} \and
Raul-Martin Rebane\inst{1}}

\authorrunning{D. Bogdanov et al.}

\institute{
Cybernetica AS, Tallinn, Estonia\\
\email{\{dan.bogdanov, eduardo.brito, annika.jaakson, peeter.laud, raul-martin.rebane\}@cyber.ee}
\and
University of Tartu, Tartu, Estonia
}

\maketitle

\begin{tikzpicture}[remember picture, overlay]
  \node[anchor=south, yshift=1cm] at (current page.south) {
    \fbox{%
      \begin{minipage}{0.9\textwidth}
        \footnotesize
        \textit{This is the extended version of the paper to appear in the Proceedings of the 5th International Workshop on Security and Privacy in Intelligent Infrastructures (SP2I 2025), held in conjunction with the 20th International Conference on Availability, Reliability and Security (ARES 2025).}
      \end{minipage}
    }
  };
\end{tikzpicture}

\begin{abstract}
This paper introduces a new set of privacy-preserving mechanisms for verifying compliance with location-based policies for vehicle taxation, or for (electric) vehicle (EV) subsidies, using Zero-Knowledge Proofs (ZKPs). We present the design and evaluation of a Zero-Knowledge Proof-of-Location (ZK-PoL) system that ensures a vehicle's adherence to territorial driving requirements without disclosing specific location data, hence maintaining user privacy. Our findings suggest a promising approach to apply ZK-PoL protocols in large-scale governmental subsidy or taxation programs.

\end{abstract}

\section{Introduction}
\label{sec:introduction}
In economic policy, taxes and subsidies are key tools to influence markets and the behaviour of economic agents. Taxes typically make activities more expensive, while subsidies reduce costs. In environmental policy, they are used to discourage activities that harm the environment and encourage those that mitigate pollution. For example, congestion and highway taxes aim to reduce vehicle traffic or fund infrastructure maintenance, while subsidies for electric vehicles (EVs) seek to lower costs and increase adoption, reducing emissions from internal combustion engines.

Implementing such schemes often requires collecting vehicle location data, which can inadvertently compromise individual privacy. Location data can easily be linked to specific drivers, raising concerns about disproportionate surveillance. Recognizing that privacy and policy enforcement need not be a zero-sum game, this paper explores novel privacy-preserving approaches to data-driven environmental policy. We are the first to (1) formalize vehicle subsidy and taxation tasks in ways that align naturally with privacy-enhancing technologies like zero-knowledge proofs (ZKPs); (2) demonstrate how existing ZKP techniques can efficiently address these tasks; and (3) analyze practical barriers to deployment, offering insights into real-world applicability of Zero-Knowledge Proof of Location (ZK-PoL) protocols.

\section{Background}
\label{sec:background}
\subsection{Location-Based Vehicle Taxes and Subsidies}
\label{sec:background:subsidies-and-taxes}

\textbf{EV subsidies.} Governments started introducing tax rebates and grants to boost electric vehicle (EV) adoption~\cite{wappelhorst2020analyzing}. Estonia’s 2019 program offered up to 5000€ (pre-tax)~\cite{estonia-ev-subsidy-2019}, with eligibility requiring:
\begin{enumerate*}
    \item driving at least 80,000 km within four years,
    \item completing 80\% of that distance in Estonia, and
    \item using renewable energy for the first 80,000 km.
\end{enumerate*}
The Environmental Investment Centre executed the program.\footnote{In 2023, the mileage requirements were relaxed to ``use the car mostly in Estonia.''} Compliance was monitored through periodic reporting or GPS tracking via third-party providers.

\textbf{Road use and emission charges.} Vehicles can be taxed via fuel sales or odometer readings~\cite{rand-rb9576}, but these methods lack road-specific granularity. Toll booths, RFID transponders, and camera systems offer precision but require expensive infrastructure. GPS tracking enables flexible, lower-cost taxation but raises privacy concerns on data storage and disclosure. Other mechanisms enforce emission standards, such as London's Ultra Low Emission Zone (ULEZ)~\cite{london-ulez}.

\textbf{Balancing verifiability and privacy.} Location-based enforcement compromises transparency, user control, and unlinkability, effectively imposing a ``privacy price.'' Sustainable mobility should not require sacrificing fundamental privacy rights. Cryptographic tolling schemes~\cite{jolfaei2023survey} and differential privacy (DP)~\cite{kim2021survey} show promise, but even obfuscated public data can leak patterns~\cite{adavoudi2024privacy}. Billing aggregation and external storage also expose users to data sovereignty risks~\cite{jolfaei2023survey}. This motivates secure computation techniques that generate verifiable proofs without disclosing raw location data, even in obfuscated form.

\subsection{Proof-of-Location and Zero-Knowledge Proofs}
\label{sec:background:proof-of-location}

Localisation techniques like GPS enable accurate position determination~\cite{obeidat2021review} but do not guarantee trust or verifiability. Proof-of-Location (PoL) extends localisation by authenticating location claims, with applications in smart mobility, content delivery, financial regulation, and smart cities~\cite{brito2025decentralized}.

PoL combines witnessing and trust~\cite{nasrulin2018robust}: a location proof is a verifiable certificate attesting a Prover’s presence at a point in space and time, supported by trusted Witness devices. Solutions combine localisation (e.g., GPS), wireless communication (e.g., Bluetooth), and cryptographic primitives (e.g., public-key cryptography). Foundational system models~\cite{brito2025decentralized} address basic PoL needs. Privacy-enhanced PoL can be achieved by layering Zero-Knowledge Proofs (ZKPs).

ZKPs allow a Prover to convince a Verifier of a statement’s validity without revealing why. Statements are typically modelled as arithmetic circuits or constraint systems defining a binary relation $R$, with a public instance $x$ and private witness $w$. A classic example is Schnorr's protocol~\cite{DBLP:conf/crypto/Schnorr89}, where knowledge of a private key is proven without revealing it. ZKPs ensure \emph{completeness} (honest proofs succeed), \emph{soundness} (cheating Provers fail), and \emph{zero-knowledge} (protocol transcripts reveal nothing beyond validity)~\cite{zk-textbook}.

For PoL applications, we adopt a centralized architecture~\cite{brito2025decentralized}, where a trusted Witness device generates signed location claims used in ZK proofs. These claims are transformed into verifiable proof statements without exposing raw trajectories~\cite{squareroots}, enabling Zero-Knowledge Proof-of-Location (ZK-PoL)~\cite{wu2020blockchain}. Our constructions specifically employ new geometric techniques to structure proofs, enabling privacy-preserving attestations of extended driving behaviour.

\section{Use cases}
\label{sec:use-cases}
\subsection{Location-based vehicle purchase subsidies}
\label{subsec:ev-problem-statement}

Our first use case for ZK-PoL protocols is proving adherence to location-based conditions for vehicle purchase subsidies. The vehicle owner acts as the Prover, and the subsidy authority as the Verifier. The goal is to prove that the vehicle has driven at least $x$ km over $T$ years, with at least $p\%$ of that distance within specified geographical bounds. For example, from Section~\ref{sec:background:subsidies-and-taxes}, $x=80\,000$, $T=4$, and $p=80$ within Estonia.

The vehicle is equipped with a trusted GPS device, assumed to be non-removable and tamper-evident. The device logs coordinates whenever the vehicle is active, and signs each trail with a private key, with the corresponding public key known to the Verifier or registered in a trusted PKI. Only the Prover has access to the raw data, transmitted via a local channel such as Bluetooth. Coordinates are projected into planar x-y values to reduce computational costs in distance calculations, using accurate projections like EPSG:3301~\cite{epsg-eesti} for Estonian territory.

After $T$ years (or periodically), the Prover generates a ZKP that the driven distance and coverage criteria are satisfied, without revealing the underlying trip coordinates. The proof verifies that the computations are based on genuine GPS data by checking the digital signatures against the known public key. The Prover can validate proof completeness before submission.

From a security perspective, preventing tampering is crucial. Without safeguards, the Prover could remove the device for trips outside the bounds or attach it to another vehicle to inflate mileage. Reliable calibration is also essential to ensure trustworthy location data. From a usability perspective, the device must be sufficiently accurate from the Prover’s standpoint. Underestimations of distance, or misclassification of trips outside the target area, could cause valid subsidy claims to fail. Dispute resolution processes for such cases are beyond the scope of this paper.

\subsection{Location-based vehicle taxation}
\label{subsec:hwtax-problem-statement}

Our second use case is proving the amount of road tax that must be paid over a given period (e.g., month or quarter). The vehicle owner acts as the Prover, and the tax authority as the Verifier. Given a taxation period $T$ and a network of toll roads, the Prover must demonstrate their driven distance $d$ on toll roads during $T$. The tax owed depends on which distance range $(x, y]$ the mileage falls into. The Prover's goal is to prove that their toll-road driving distance satisfies $d \leq y$ for a selected bracket, minimizing their tax liability.

The setup mirrors the subsidy case: a trusted, tamper-evident GPS device is attached to the vehicle, logging and signing location data. At the end of each taxation period, the Prover generates a ZKP proving their distance falls within a chosen bracket without revealing trip details, and submits it to the Verifier. Proof completeness ensures the Prover can select the lowest eligible tax bracket.

Unlike the subsidy case, proof submission is mandatory each period, with penalties for non-submission. Trust in device attachment and calibration is critical to prevent fraud, such as detaching the device or leaving it stationary. As before, device accuracy is essential, though dispute resolution mechanisms are outside the scope of this paper.

\section{System Model}
\label{sec:system_model}
\begin{figure}[ht]
    \centering
    \includegraphics[width=0.5\columnwidth]{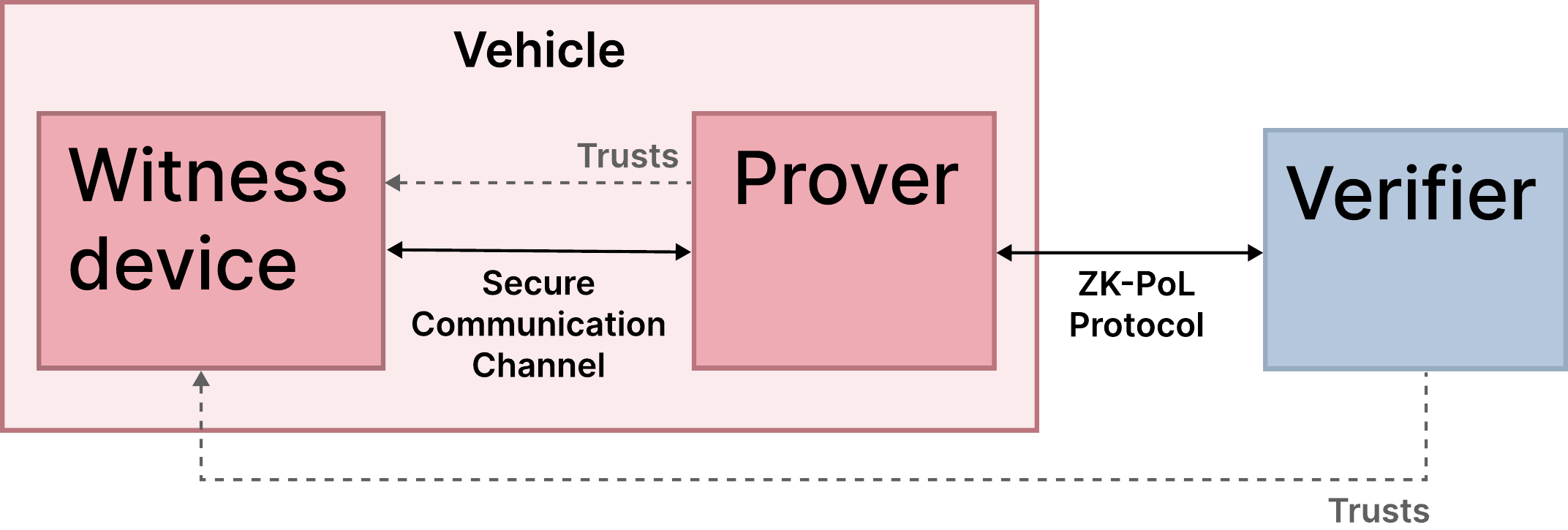}
    \caption{ZK-PoL system model for mobility use cases.}
    \label{fig:system_model}
\end{figure}

Building on these mobility use cases, we now instantiate the ZK-PoL model, noting its applicability to other domains such as asset or personnel tracking. The system model (Figure~\ref{fig:system_model}) centres on three components: the Witness device, the Prover, and the Verifier.

The Witness device is a tamper-resistant or tamper-evident device coupled to the vehicle. Using its GPS module, it generates raw location claims, and signs them, without interpreting eligibility rules or computing taxes, allowing it to remain update-free even if policy changes. Trust is established between the Witness device and both the Prover and the Verifier, who rely on it to produce continuous, precise, and verifiable location data.

The Prover is any entity (owner, driver, or custodian) responsible for demonstrating location compliance. They establish a restricted communication channel with the Witness device to receive signed location data. Our model is not tied to proximity constraints or specific communication ranges, unlike some related PoL work identified in Section~\ref{sec:background:proof-of-location}. Importantly, proofs are bound to the vehicle's identity via the Witness device, not the Prover’s personal identity. The Prover assembles the proof and manages subsequent interactions with the Verifier, including executing the ZKPs.

The Verifier trusts the Witness device's public key $\mathit{pk}$, used to sign the location claims. The Prover’s proof asserts that a trajectory exists, signed and verifiable under $\mathit{pk}$, and satisfies the conditions for subsidy eligibility or applicable tax rates.

\section{Zero-Knowledge Protocol Specification}
\label{sec:zk-specification}
Encoding ZK statements directly into low-level paradigms is difficult and error-prone, as these systems offer minimal abstraction and require bit-level reasoning. Higher-level approaches address this gap, including gadget APIs~\cite{circom}, DSLs~\cite{snarkl,xjsnark,viaduct}, and general-purpose languages~\cite{geppetto}. We use \ZKSC~\cite{zksc-CSF24,zksc-arxiv}, a high-level imperative DSL with C++/Rust-like syntax, designed specifically for ZK proofs. It features an information flow type system~\cite{volpanosmithirvine,jflow} that separates Prover-private data and shared data, enforcing confidentiality and enabling efficient local computations. It also distinguishes compile-time and runtime values, supporting protocols that require relation preprocessing~\cite{groth16}. Local values injected into the proof must pass correctness checks to maintain soundness~\cite{settyetal,buffet}. \ZKSC offers a standard library~\cite{zksecrec-github} of optimized primitives, including bit operations, fixed-point arithmetic~\cite{fixedpoints,squareroots}, and efficient constructions for RAM and inequality proofs~\cite{arya,buffet}. The compiler targets formats like SIEVE~IR~\cite{sieveir} and integrates with interactive ZKP systems~\cite{macncheese,EMP} via a Rust trait~\cite{rustbook-traits}, abstracting circuit construction through gates, constraints, and inputs.

\begin{figure}[htbp]
    \centering
    \begin{subfigure}[b]{0.42\columnwidth}
        \centering
        \includegraphics[width=\linewidth]{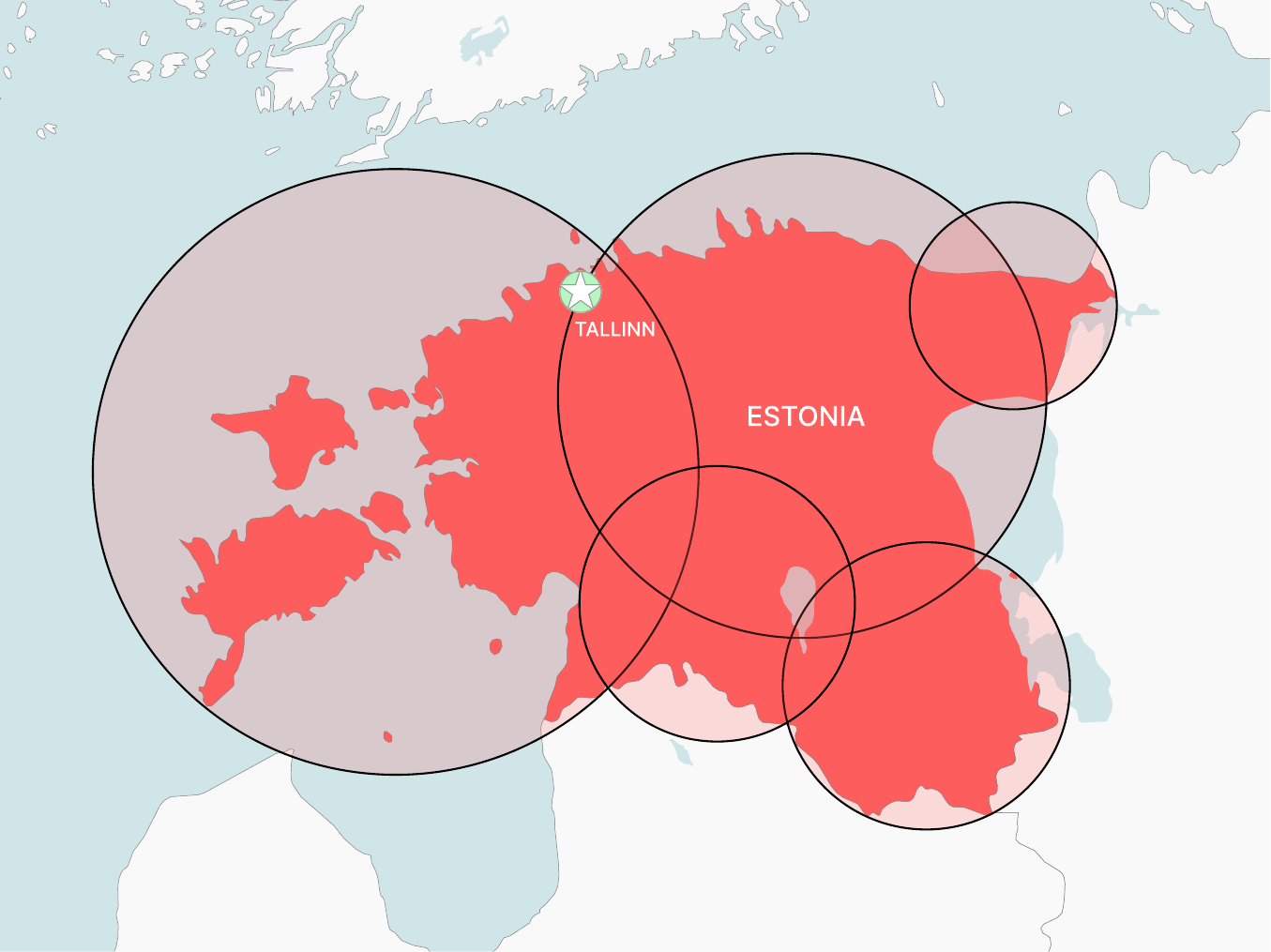}
        \caption{Approximation of Estonia by circles.}
        \label{fig:circles_territory}
    \end{subfigure}
    \hfill
    \begin{subfigure}[b]{0.42\columnwidth}
        \centering
        \includegraphics[width=\linewidth]{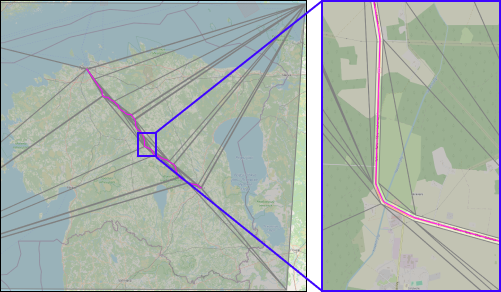}
        \caption{Triangles covering Tallinn–Tartu road.}
        \label{fig:hwtax_triangles}
    \end{subfigure}
    \caption{Geographical approximations used for the EV subsidy (left) and highway tax (right) use cases.}
    \label{fig:geo_approximations}
\end{figure}

\subsection{EV subsidy use-case}
\label{subsec:ev-protocol-description}

The ZKP protocol goal in the EV subsidy case is for the Prover to demonstrate, with a list of coordinate points, that the total trail length exceeds a required distance, and a required percentage falls within specified geographical bounds.

\begin{figure}[htbp]
    \centering
    \begin{subfigure}[b]{0.42\linewidth}
        \centering
        \includegraphics[width=\linewidth]{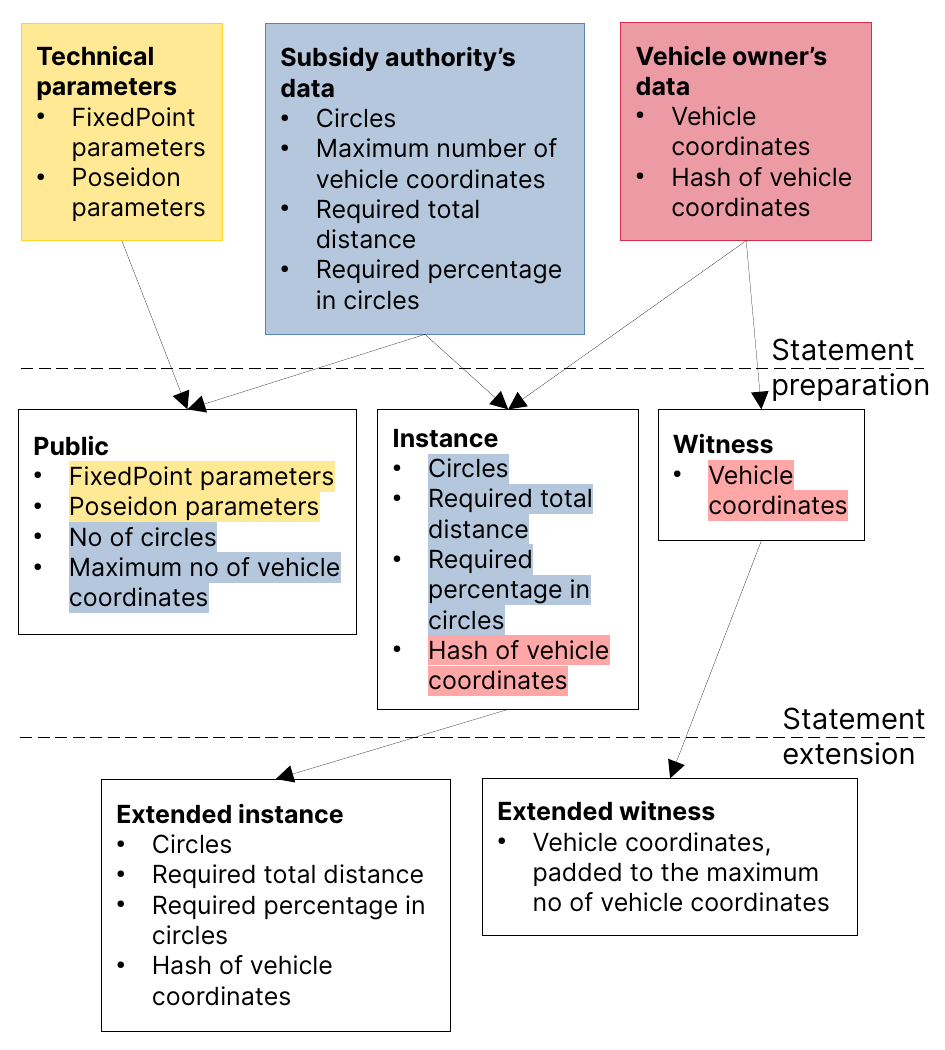}
        \caption{EV subsidy use case.}
        \label{fig:ev_data_model}
    \end{subfigure}
    \hfill
    \begin{subfigure}[b]{0.42\linewidth}
        \centering
        \includegraphics[width=\linewidth]{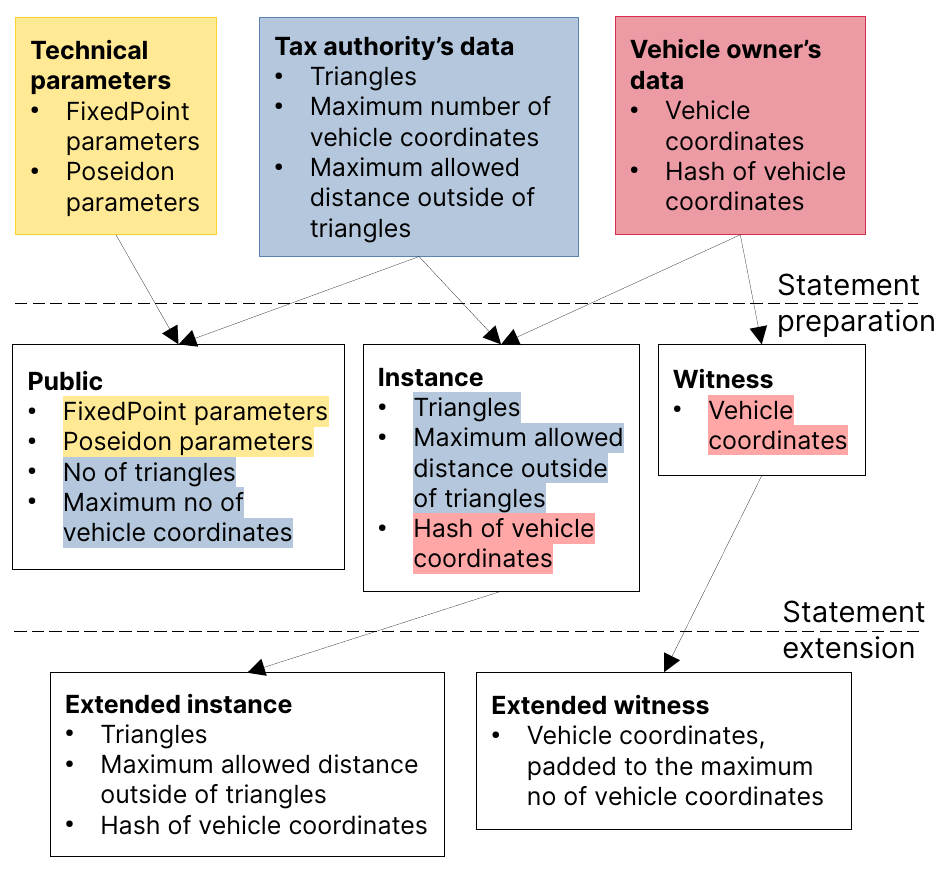}
        \caption{Highway tax use case.}
        \label{fig:hwtax_data_model}
    \end{subfigure}
    \caption{Data models for the EV subsidy (left) and highway tax (right) use cases.}
    \label{fig:data_models}
\end{figure}

We use a planar coordinate system to avoid expensive trigonometric operations. Geographical bounds are approximated by a union of circles (Figure~\ref{fig:circles_territory}), which flexibly balances precision and computational cost. The proof data structure is shown in Figure~\ref{fig:ev_data_model}. Besides the technical parameters, discussed later, there are the values characterizing the \emph{size} of the statement. Certain sizes have to be fixed; the running time of the protocol will depend (only) on these sizes.  To simplify proofs, we assume:
\begin{itemize}
    \item all coordinates and radii are positive integers,
    \item the trajectory has at most a known maximum number of points,
    \item the coordinate list is padded and hashed to hide the actual trail length.
\end{itemize}

\begin{algorithm}[ht]
\caption{Proof for the electric vehicle use-case.}\label{alg:evsubsidy}
\KwIn{Parameters $\prepub{\mathit{pp}}$ for Poseidon hash function}
\KwIn{Centers $(\prever{u_i},\prever{v_i})$ and radii $\prever{r_i}$ of circles ($1\leq i\leq \prepub{n_\mathrm{circ}}$)}
\KwIn{Required distance $\prever{d_\mathrm{req}}$ and percentage $\prever{P_\mathrm{req}}$}
\KwIn{Coordinates $(\prepro{x_i},\prepro{y_i})$ on the trajectory ($1\leq i\leq \prepub{n_\mathrm{traj}}$)}
\KwIn{Expected hash $\prever{h_\mathrm{ex}}$ of the list of coordinates}
$\postpro{\mathit{tot}} \leftarrow 0$,
$\postpro{\mathit{cc}} \leftarrow 0$\\
$\postpro{\vec x}\leftarrow \wire{\prepro{\vec x}}$, 
$\postpro{\vec y}\leftarrow \wire{\prepro{\vec y}}$\\ 
$\assert{\mathsf{hash}(\prepub{\mathit{pp}},\postpro{\vec x} \| \postpro{\vec y}) = \wire{\prever{h_\mathrm{ex}}}}$\\
$\postver{\vec u}\leftarrow \wire{\prever{\vec u}}$, 
$\postver{\vec v}\leftarrow \wire{\prever{\vec v}}$\\
\lFor{$i\leftarrow1$ \KwTo $\prepub{n_\mathrm{circ}}$}{$\postver{s_i}\leftarrow\wire{\prever{r_i}^2}$}
$\postpro{\mathit{b\_pi}}\leftarrow \mathsf{check\_inside}(\postver{\vec u}, \postver{\vec v}, \postver{\vec s}, \postpro{x_1},\postpro{y_1})$\\ 
\For{$i\leftarrow2$ \KwTo $\prepub{n_\mathrm{traj}}$}{
    $\postpro{\mathit{b\_in}}\leftarrow\mathsf{check\_inside}(\postver{\vec u}, \postver{\vec v}, \postver{\vec s}, \postpro{x_i},\postpro{y_i})$\\
    $\postpro{d_i}\leftarrow\mathsf{sqrt}((\postpro{x_i}-\postpro{x_{i-1}})^2+(\postpro{y_i}-\postpro{y_{i-1}})^2)$\\
    $\postpro{\mathit{tot}} := \postpro{\mathit{tot}} + \postpro{d_i}$\\
    $\postpro{\mathit{cc}} := (\postpro{\mathit{b\_pi}}\wedge\postpro{\mathit{b\_in}}) \mathbin{?} (\postpro{\mathit{cc}} + \postpro{d_i}) \mathbin{:} \postpro{\mathit{cc}}$\\
    $\postpro{\mathit{b\_pi}} := \postpro{\mathit{b\_in}}$
}
$\assert{\wire{\prever{d_\mathrm{req}}}\leq\postpro{\mathit{tot}}}$\\
$\assert{\postpro{\mathit{tot}} * \wire{\prever{P_\mathrm{req}}} \leq \postpro{\mathit{cc}} * 100}$
\end{algorithm}

\begin{algorithm}
\caption{Checking that a point is inside a set of circles: $\mathsf{check\_inside}$.}\label{alg:checkinside}
\KwIn{Centers $(\postver{u_i},\postver{v_i})$ and squares of radii $\postver{s_i}$ of circles ($1\leq i\leq \prepub{n_\mathrm{circ}}$)}
\KwIn{Coordinates $\postpro{x},\postpro{y}$ of a point}
\KwOut{Boolean $\postpro{b}$ indicating whether the point is inside at least one of the circles}
$\postpro{b}\leftarrow\mathsf{false}$\\
\For{$i\leftarrow1$ \KwTo $\prepub{n_\mathrm{circ}}$}{
    $\postpro{b} := \postpro{b} \vee \Bigl((\postpro{x}-\postver{u_i})^2 + (\postpro{y}-\postver{v_i})^2 \leq \postver{s_i}\Bigr)$
}
\Return{$\postpro{b}$}
\end{algorithm}

We presume that revealing the hash $h_\mathrm{ex}$ of the coordinate trail to the Verifier does not significantly compromise privacy. This hash, signed by the Witness device and transmitted via the Prover, binds the ZK proof to the Witness’s public key, enabling the Verifier to independently verify authenticity in parallel to proof verification.

Algorithm~\ref{alg:evsubsidy} checks that the coordinate trail satisfies the subsidy conditions, using a subroutine from Algorithm~\ref{alg:checkinside} to determine circle inclusion. In our pseudocode, values managed by the ZK protocol are enclosed in double brackets, while locally managed values are not. Color coding indicates visibility: \prdom{red} for Prover-only, \vedom{green} for shared data, and \pudom{blue} for public constants. Arithmetic and comparisons on ZK-managed values are realized securely by the underlying protocol, ensuring that cheating by the Prover is infeasible. Our conventions extend the notation commonly used for privacy-preserving computations, where one typically uses square brackets (either single or double) around a variable to denote that the value of this variable is handled by the cryptographic technology.

Variable types are not made explicit in the pseudocode but are deduced as follows: variables prefixed with $b$ are booleans, $h_\mathrm{ex}$ is a hash output, $\mathit{pp}$ contains hash parameters, and the rest are integers. All necessary primitives, including integer square roots (rounded toward zero), are provided by the \ZKSC standard library~\cite{zksecrec-github}. For the use of the square root, see also the discussion in Sec.~\ref{ssec:optimizations}.

We use the Poseidon hash function~\cite{grassi2021poseidon}, optimized for field operations in ZKP systems, instead of general-purpose bit-based hashes like SHA-256, which would incur costly bit extraction and manipulation overheads. Poseidon parameters are precomputed and included as public parameters.

In Algorithm~\ref{alg:evsubsidy}, $\mathsf{wire}$ marks values as ZK protocol inputs, and $\mathsf{assert}$ enforces conditions during proof execution. If a $\mathsf{wire}$ argument is known to both parties, its value is trusted; if private to the Prover, it requires independent checks within the proof.

ZK protocols cannot branch on conditions that are also computed under ZK or depend on private values. Therefore, updating the distance in circles $\mathit{cc}$ is non-trivial. We implement it using an \emph{oblivious choice}, a ternary operation $b\mathbin{?}x\mathbin{:}y$ that returns $x$ if $b$ is true and $y$ otherwise. For integers $b,x,y$ with $b\in\{0,1\}$, this can be computed as $y + b(x - y)$.

The proof statement structure (see Alg.~\ref{alg:evsubsidy}) proceeds as follows. First, the Prover hashes the coordinate trail (padded with the last point) and asserts that it matches the expected instance hash. Then, it computes the total trail length (\texttt{tot}) and the length of segments where both endpoints lie within some circle (\texttt{cc}). Finally, it asserts:
\begin{align*}
\texttt{tot} \geq d_{req} && \text{and} && \texttt{cc} \geq \texttt{tot} \cdot \frac{P_{req}}{100}
\end{align*}

If both assertions hold, the coordinate trail satisfies the subsidy rules. To compute \texttt{tot} and \texttt{cc} (see Alg.~\ref{alg:evsubsidy}), we evaluate the Euclidean distance
\[
d_i = \sqrt{(x_i - x_{i-1})^2 + (y_i - y_{i-1})^2}
\]
for each pair of consecutive points, and check whether both points lie within some circle (not necessarily the same one). The inclusion check for a point $(x_i, y_i)$ in a circle with centre $(u_j, v_j)$ and radius $r_j$ (see Alg.~\ref{alg:checkinside}) is performed via:
\begin{align}
    (x_i - u_j)^2 + (y_i - v_j)^2 \leq r_j^2 \label{eq:in_circle_check}
\end{align}
This comparison is done for each point and all circles $j \in \{0, \ldots, \texttt{$n_{traj}$}\}$, accepting if any match. The values \texttt{tot} and \texttt{cc} are then:
\begin{align*}
    \texttt{tot} = \sum_{i \in T} d_i \quad\quad \texttt{cc} = \sum_{i \in S} d_i
\end{align*}
where $S \subset T$ includes indices $i$ where both endpoints lie in some circle. Square roots are required to compute distances but are not needed for inclusion checks, which use squared distances and the inequality \eqref{eq:in_circle_check}, thus saving computation under ZK.

\subsection{Highway tax use-case}
\label{subsec:hwtax-protocol-description}

The goal of the ZKP statement in the highway tax use-case is to prove that no more than a specified distance of the Prover’s coordinate trail lies along taxed roads (see Section~\ref{subsec:hwtax-problem-statement}). As in the subsidy use-case, an approximate representation is required. However, while subsidy proofs aim to show that points are \emph{inside} a designated area, here the Prover seeks to prove points are \emph{outside} taxed roads. Thus, we represent the untaxed region instead by triangulating the area surrounding the highways. Being outside the taxed roads becomes equivalent to being inside one of these triangles.

Figure~\ref{fig:hwtax_triangles} illustrates this for the road between Tallinn and Tartu in Estonia: the road is marked in pink, and the surrounding area (including all of Estonia) is shown as gray triangles. A buffer margin around the road accounts for GPS noise and controls the precision-performance trade-off: tighter margins require more segments and thus more triangles, increasing proof cost.

The structure of the proof data is shown in Figure~\ref{fig:hwtax_data_model}. We adopt the same input constraints as in the EV use-case (Section~\ref{subsec:ev-protocol-description}), with one additional requirement: all points in the trail must either lie near the highway or within the public triangulation, i.e., within Estonian territory as depicted in Figure~\ref{fig:hwtax_triangles}.

\begin{algorithm}[ht]
\caption{Proof for the highway tax use case.}\label{alg:hwtax}
\KwIn{Parameters $\prepub{\mathit{pp}}$ for Poseidon hash function}
\KwIn{Vertices $(\prever{X_{j,k}}, \prever{Y_{j,k}})$ of triangles ($1\leq j\leq \prepub{n_\mathrm{tri}}$, $k \in \{1, 2, 3\}$)}
\KwIn{Maximum allowed on-highway distance $\prever{d_\mathrm{max}}$}
\KwIn{Coordinates $(\prepro{x_i},\prepro{y_i})$ on the trajectory ($1\leq i\leq \prepub{n_\mathrm{traj}}$)}
\KwIn{Expected hash $\prever{h_\mathrm{ex}}$ of the list of coordinates}

$\postpro{\mathit{tot}} \leftarrow 0$, $\postpro{\mathit{hw}} \leftarrow 0$\\
$\postpro{\vec x}\leftarrow \wire{\prepro{\vec x}}$, $\postpro{\vec y}\leftarrow \wire{\prepro{\vec y}}$\\
$\assert{\mathsf{hash}(\prepub{\mathit{pp}}, \postpro{\vec x}\| \postpro{\vec y}) = \wire{\prever{h_\mathrm{ex}}}}$\\
$\postver{X}\leftarrow \wire{\prever{X}}$, $\postver{Y}\leftarrow \wire{\prever{Y}}$\\

\For{$i\leftarrow1$ \KwTo $\prepub{n_\mathrm{traj}}$}{
    $\postpro{t_i}\leftarrow \wire{\mathsf{find\_triangle}(\prepro{x_i},\prepro{y_i},\prever{X},\prever{Y})}$\\
    $\postpro{{\vec a}_i} \leftarrow \mathsf{lookup}(\postpro{t_i},\postver{X})$ \tcp*{$|{\vec a}_i|=3$}
    $\postpro{{\vec b}_i} \leftarrow \mathsf{lookup}(\postpro{t_i},\postver{Y})$\\
    $\postpro{c_i} \leftarrow \mathsf{check\_inside\_triangle}(
    \postpro{{\vec a}_i}, \postpro{{\vec b}_i}, \postpro{x_i}, \postpro{y_i})$\\
    \If{$i > 1$}{
        $\postpro{d_i}\leftarrow\mathsf{sqrt}((\postpro{x_i}-\postpro{x_{i-1}})^2+(\postpro{y_i}-\postpro{y_{i-1}})^2)$\\
        $\postpro{\mathit{tot}} \leftarrow \postpro{\mathit{tot}} + \postpro{d_i}$\\
        $\postpro{\mathit{hw}} \leftarrow (\postpro{\mathit{c_{i-1}}} \wedge \postpro{\mathit{c_i}}) \mathbin{?} (\postpro{\mathit{hw}} + \postpro{d_i}) \mathbin{:} \postpro{\mathit{hw}}$\\
    }
}
$\assert{\postpro{\mathit{tot}} - \postpro{\mathit{hw}} \leq \wire{\prever{d_\mathrm{max}}}}$
\end{algorithm}

\begin{algorithm}
\caption{Checking whether a point is inside a triangle: $\mathsf{check\_inside\_triangle}$.}\label{alg:checkintriangle}
\KwIn{Vertices $(\postpro{a_1}, \postpro{b_1})$, $(\postpro{a_2}, \postpro{b_2})$, $(\postpro{a_3}, \postpro{b_3})$ of a triangle}
\KwIn{Coordinates $\postpro{x}, \postpro{y}$ of a point}
\KwOut{Boolean indicating whether the point is inside the triangle}

$\postpro{A}\leftarrow\Delta\mathsf{area\_dbl}(\postpro{a_1},\postpro{b_1},\postpro{a_2},\postpro{b_2},\postpro{a_2},\postpro{b_3})$\\
$(\prepro{s},\prepro{t})\leftarrow\mathsf{get\_bcoords}(\prepro{x},\prepro{y},\prepro{a_1},\prepro{b_1},\prepro{a_2},\prepro{b_2},\prepro{a_3},\prepro{b_3})$\\
$\postpro{s}\leftarrow\wire{\prepro{s}}$, $\postpro{t}\leftarrow\wire{\prepro{t}}$, $\postpro{u}\leftarrow\postpro{A}-\postpro{s}-\postpro{t}$\\
$\postpro{x'}\leftarrow \postpro{u}\cdot\postpro{a_1}+\postpro{s}\cdot\postpro{a_2}+\postpro{t}\cdot\postpro{a_3}$\\
$\postpro{y'}\leftarrow \postpro{u}\cdot\postpro{b_1}+\postpro{s}\cdot\postpro{b_2}+\postpro{t}\cdot\postpro{b_3}$\\
$\assert{ \postpro{x'} = \postpro{x}\cdot\postpro{A} \wedge \postpro{y'} = \postpro{y}\cdot\postpro{A}}$\\
\Return{$0\leq \postpro{s} \wedge 0\leq\postpro{t} \wedge 0 \leq \postpro{u}$}
\end{algorithm}

The pseudocode for the ZKP statement is given in Alg.~\ref{alg:hwtax}, with a helper routine in Alg.~\ref{alg:checkintriangle}. The conventions follow those in Alg.~\ref{alg:evsubsidy}. The statement is more complex here, as checking triangle inclusion is more involved than for circles. A significant portion of computation --- functions $\mathsf{find\_triangle}$ and $\mathsf{get\_bcoords}$ --- is performed locally by the Prover. The results of these local computations are verified on the circuit, and the details of these computations do not affect the validity of the proof. The implementation details of $\mathsf{find\_triangle}$ and $\mathsf{get\_bcoords}$ are discussed in App.~\ref{app:hwtax}.

As in Alg.~\ref{alg:evsubsidy}, the ZK protocol begins by inputting the coordinate trail and asserting that its hash matches the one in the instance. It then computes the total trail length (\texttt{tot}) and the portion inside the triangle set (\texttt{hw}). Finally, it asserts that $\texttt{tot} - \texttt{hw} \leq d_\mathrm{max}$. If this check passes, the Prover is deemed to have driven at most \texttt{$d_{max}$} on taxed roads.

Alg.~\ref{alg:hwtax} differs from Alg.~\ref{alg:evsubsidy} in the logic of checking whether a point is inside a region. While in Alg.~\ref{alg:evsubsidy} we simply performed the test for all circles and took the disjunction of results, in Alg.~\ref{alg:hwtax} Prover obliviously picks a triangle and states that the point belongs to this particular triangle. The triangle-finding function $\mathsf{find\_triangle}$ gives the index of that triangle (or an arbitrary index if the point does not belong to any triangle); this index $t_i$ becomes another input to the ZK protocol. Next, we use the techniques of Oblivious RAM under ZK to locate the coordinates of the vertices of the chosen triangle; the function $\mathsf{lookup}$ gives us the $x$- and $y$-coordinates of this particular triangle. The function $\mathsf{lookup}$ can be implemented in various ways; our implementation (shown in Alg.~\ref{alg:oram}) is based on expanding $t_i$ to its characteristic vector and computing its scalar product with the vector of coordinates.

\begin{algorithm}
\caption{Looking up the coordinates of vertices: $\mathsf{lookup}$.}\label{alg:oram}
\KwIn{Index $\postpro{t}$}
\KwIn{Single coordinates of vertices of triangles $\postver{X_{j,k}}$ ($1\leq j\leq \prepub{n_\mathrm{tri}}$, $k \in \{1, 2, 3\}$)}
\KwOut{Single coordinates $\postpro{a_k}$ of the selected triangle ($k \in \{1, 2, 3\}$)}
$\postpro{s}\leftarrow 0$\\
\For{$i\leftarrow 1$ \KwTo $\prepub{n_\mathrm{tri}}$}{
    \leIf{$i=\prepro{t}$}{$\prepro{x_i}\leftarrow 1$}{$\prepro{x_i}\leftarrow 0$}
    $\postpro{x_i}\leftarrow\wire{\prepro{x_i}}$\\
    $\assert{\postpro{x_i} = 0 \vee \postpro{x_i} = 1}$\\
    $\postpro{s} := \postpro{s} + \postpro{x_i}$
}
$\assert{\postpro{s} = 1}$\\
$\postpro{a_1}\leftarrow 0$, $\postpro{a_2}\leftarrow 0$, $\postpro{a_3}\leftarrow 0$\\
\For{$i\leftarrow 1$ \KwTo $\prepub{n_\mathrm{tri}}$}{
    \lFor{$k\leftarrow1$ \KwTo $3$}{$\postpro{a_k}:=\postpro{a_k}+\postpro{x_i}\cdot\postver{X_{i,k}}$}
}
\Return{$\postpro{a_1},\postpro{a_2},\postpro{a_3}$}
\end{algorithm}

To check whether a point $(x, y)$ lies inside a triangle $[(a_1, b_1), (a_2, b_2), (a_3, b_3)]$ in ZK (Alg.~\ref{alg:checkintriangle}), we use its \emph{barycentric coordinates\footnote{\url{https://en.wikipedia.org/wiki/Barycentric\_coordinate\_system}}} $(s, t, u)$ with respect to the triangle’s vertices. We use unnormalized coordinates, where $s + t + u = A$ for all valid coordinate triples and for some positive constant $A$. These coordinates are computed locally by the Prover via $\mathsf{get\_bcoords}$, using only local inputs. The function $\mathsf{check\_inside\_triangle}$ then reconstructs the Cartesian point from $(s, t, u)$ and asserts that it equals $(x, y)$. It returns true if and only if $s, t, u \geq 0$, meaning the point lies inside or on the boundary of the triangle.

We choose $A$ as twice the area of the triangle $[(a_1, b_1), (a_2, b_2), (a_3, b_3)]$, which allows $(s, t, u)$ to be computed without division. It is given by

\begin{equation}\label{eq:dblarea}
\mathsf{area\_dbl}(a_1,b_1,a_2,b_2,a_3,b_3)=\left|\mathrm{det}\left(\begin{matrix}
        a_1 & a_2 & a_3 \\
        b_1 & b_2 & b_3 \\
        1 & 1 & 1
\end{matrix} \right)\,\right|
\end{equation}

\subsection{Possible optimizations}\label{ssec:optimizations}

Several checks in our protocols can be optimized based on the Prover’s goals. For instance, Alg.~\ref{alg:hwtax} allows a relaxed behavior: when locating the triangle containing a point $(x_i, y_i)$, any incorrect index selection only disadvantages the Prover. If $\mathsf{check\_inside\_triangle}$ fails, the point is counted as outside, increasing the taxable distance. In both Alg.~\ref{alg:evsubsidy} and~\ref{alg:hwtax}, the distance $d_i$ between consecutive points is computed by ensuring
\[
d_i^2 \leq (x_i-x_{i-1})^2+(y_i-y_{i-1})^2<(d_i+1)^2,
\]
Here, $d_i$ is input by the Prover, and the ZK protocol checks the inequalities. Depending on the application, one of the checks may be omitted to reduce circuit size: in the highway tax use-case, where minimizing distance is the goal, only the upper bound matters; in the subsidy use-case, where demonstrating sufficient travel distance is important, only the lower bound is essential. If subsidy eligibility depended solely on distance inside circles, omitting the upper bound check would also be valid.

\subsection{Security and Privacy Analysis}

We summarize how our protocols fulfil key security and privacy requirements, while maintaining core cryptographic properties. A formal proof sketch is provided in App.~\ref{app:security}.
Soundness, completeness, and zero-knowledge are guaranteed by:
\begin{itemize}
\item Correct cryptographic implementation of the underlying ZKP back-ends (e.g., \texttt{emp-zk}~\cite{emp-zk-github}, Diet Mac'n'Cheese~\cite{galois-swanky-github}),
\item Correctness of the proof algorithms (Alg.~\ref{alg:evsubsidy}--\ref{alg:oram}),
\item Enforcement of information flow and visibility policies by the \ZKSC type system~\cite[Thm.~3--5]{zksc-CSF24},
\item Secure linkage of location data via the tamper-evident Witness device.
\end{itemize}

ZK-SecreC enables high-level construction of ZKPs while enforcing data visibility and input validation policies, critical to preserving privacy guarantees. The compiled circuits target interactive ZK backends, such as \texttt{emp-zk} and Diet Mac’n’Cheese, where proof generation proceeds through multiple rounds of communication rather than producing a single static proof artifact. Correctness at both the high-level and back-end levels is essential, as errors—e.g., in fixed-point arithmetic or unchecked values—could undermine soundness or zero-knowledge.

Our trust model assumes that the Witness device securely signs location data, and that raw GPS trails remain private to the Prover. The Verifier learns only whether compliance conditions are met, via privacy-preserving proofs. Unlike conventional systems (see Section~\ref{sec:background}) that expose full location logs or odometer readings, our approach preserves data and computational sovereignty while revealing no more than necessary. Compared to methods like differential privacy (DP) or secure multi-party computation (MPC), ZKPs enable individual compliance proofs without coordination~\cite{10.1007/978-3-662-47854-7_14} or statistical leakage~\cite{jolfaei2023survey}.

\section{Performance Evaluation}
\label{sec:performance_evaluation}
We prototyped both use cases to assess whether their performance with current ZKP technology is acceptable for real-world deployment.
We used a Raspberry Pi 3B/4 with an embedded GPS module in place of the Witness device (in an actual deployment, this device should be tamper-resistant). The device contained a private signing key corresponding to a known public key.
It captured coordinates every 30 seconds, consistent with commercial fleet tracking standards~\cite{verizon-fleet-tracker}. Test input sizes (200, 3600, and 43,800 points) correspond to typical single trips, monthly, and yearly driving durations~\cite{american-driving-survey-22}.

In the EV subsidy use case, we approximated Estonia's territory with 5 circles; in the highway-tax use case, the area along a 179 km highway (Tallinn–Tartu) was modelled using 248 triangles, achieving a maximal deviation of 50 meters from the real road. The number of shapes depends on the geographic area and tolerated error.

\begin{figure}[ht]
    \centering
    \begin{subfigure}[t]{0.42\columnwidth}
        \centering
        \includegraphics[width=\linewidth]{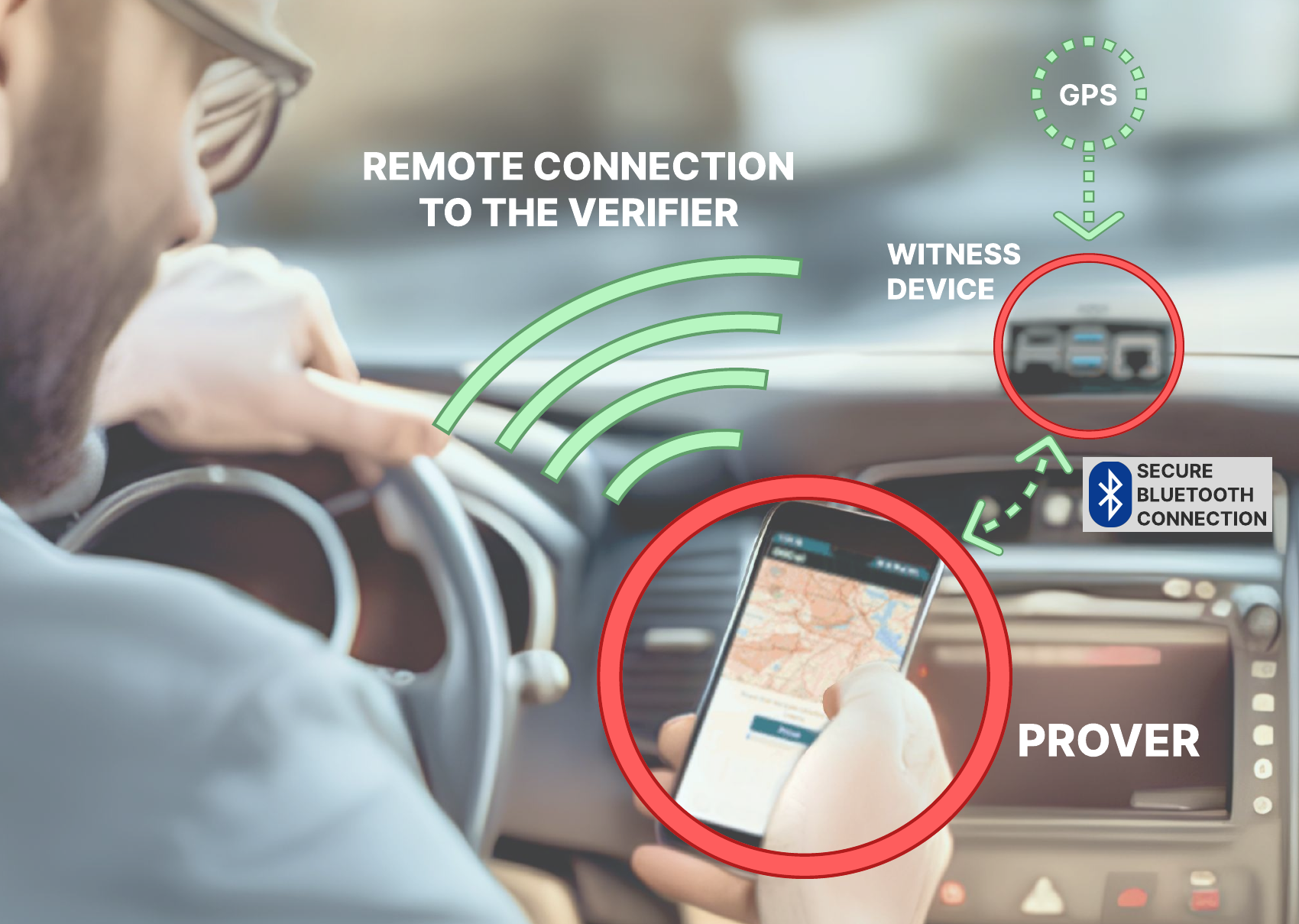}
        \caption{Prototype setup: Android phone (Prover) and Raspberry Pi (Witness) connected via Bluetooth.}
        \label{fig:real_world_prototype}
    \end{subfigure}
    \hfill
    \begin{subfigure}[t]{0.55\columnwidth}
        \centering
        \includegraphics[width=\linewidth]{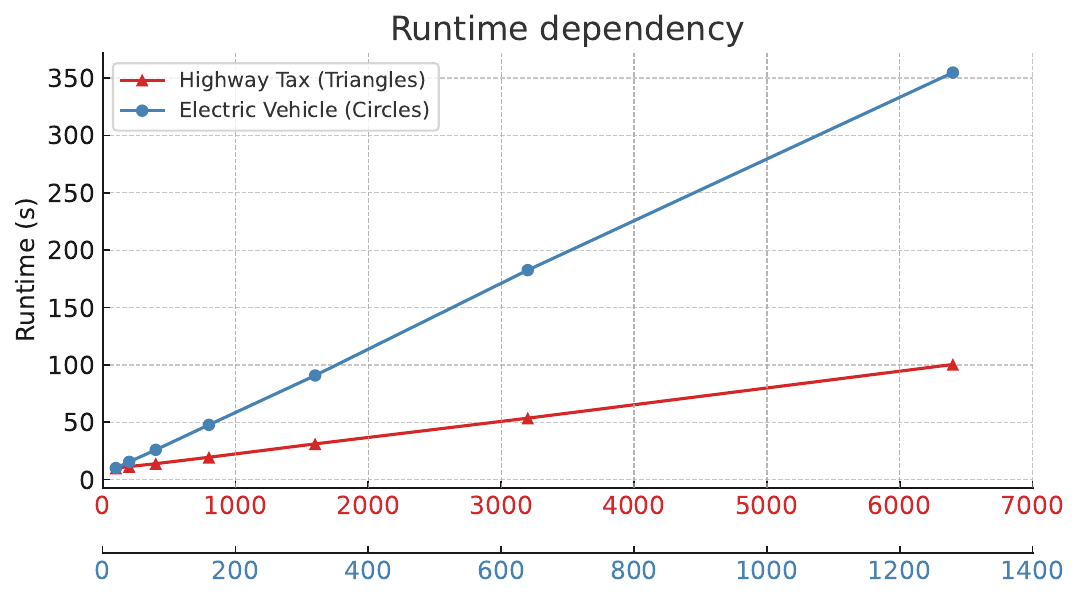}
        \caption{Runtime vs. number of circles (EV) or triangles (Highway Tax) for area approximation.}
        \label{fig:ev-perf-circles}
    \end{subfigure}
    \caption{(a) Real-world prototype; (b) Local runtime performance.}
    \label{fig:prototype-and-runtime}
\end{figure}

The Raspberry Pi sent signed location data over a secure Bluetooth connection to a Prover device.
A native app handled Bluetooth configuration, loaded a WebAssembly-based executable via WebView, and executed the ZKP protocol with a remote Verifier server. Figure~\ref{fig:real_world_prototype} shows the real-world prototype setup. Outside the browser, we tested emp-zk~\cite{emp-zk-github} and Diet Mac'n'Cheese (MnC)~\cite{galois-swanky-github} back-ends, though comparisons are slightly biased due to better MnC integration. Emp-zk tests were run locally (0 ms delay). MnC tests were run under three network conditions: Fast (10 ms, 300 Mbit/s), Medium (80 ms, 100 Mbit/s), and Slow (200 ms, 10 Mbit/s). Verifiers used a SuperServer 6028R-TR (2×Intel Xeon E5-2640 v3, 128GB RAM), while Provers included a Lenovo ThinkPad T14s Gen 1 (Ryzen 7 PRO 4750U), a Google Pixel 5a, and a Raspberry Pi Compute Module 4. Details appear in Table~\ref{tab:bench_results}. Due to differing protocol parameters, performance sensitivity to latency varies across devices.

We also evaluated how performance scales with the number of circles or triangles. Using single-trip input size, we varied the number of shapes and observed runtime trends (Figure~\ref{fig:ev-perf-circles}). For this test, the PC Prover ran both Prover and Verifier roles without network delay, isolating computational complexity. Results confirm a linear dependency on the number of shapes. For large shape counts, the highway-tax algorithm (Alg.~\ref{alg:hwtax}) outperforms the EV subsidy algorithm (Alg.~\ref{alg:evsubsidy}) in point-in-shape checking, due to use-case-specific optimizations. These trends are expected to hold for longer vehicle trajectories due to the linear complexity of the underlying algorithms (Section~\ref{sec:zk-specification}).

{\renewcommand{\arraystretch}{1.1}
\begin{table*}
\scriptsize
\centering
\begin{tabular}{l|l|l|l|l|l|l|}
\cline{2-7}
                               & Pi 4 (F)       & Pi 4 (M)      & Pi 4 (S)      & Phone (F)      & Phone (M)     & Phone (S)      \\ \hline
\multicolumn{1}{|l|}{EV trip}  & 12.01 (0.15)   & 17.49 (0.09)  & 29.67 (0.20)  & 13.60 (0.15)   & 21.02 (0.44)  & 32.84 (0.07)   \\ \hline
\multicolumn{1}{|l|}{EV month} & 47.28 (0.33)   & 54.12 (0.20)  & 69.61 (0.44)  & 127.88 (0.88)  & 151.92 (2.49) & 189.2 (1.90)   \\ \hline
\multicolumn{1}{|l|}{EV year}  & 1229.03 (17.07) & 1275.28 (15.18) & 1309.39 (21.75) & 1724 (16.43)   & 2078 (26.74)  & 2371 (27.93)   \\ \hline
\multicolumn{1}{|l|}{HW trip}  & 25.57 (0.15)   & 31.52 (0.21)  & 45.72 (0.58)  & 36.44 (0.25)   & 43.98 (0.35)  & 54.33 (0.49)   \\ \hline
\multicolumn{1}{|l|}{HW month} & 323.55 (4.96)  & 336.82 (2.44) & 387.70 (4.92) & 369.39 (1.44)  & 388.72 (5.88) & 403.26 (6.08)  \\ \hline
\multicolumn{1}{|l|}{HW year}  & -              & -             & -             & -              & -             & -              \\ \hline
\end{tabular}

\hfill

\begin{tabular}{l|l|l|l|l|}
\cline{2-5}
                               & PC (F)         & PC (M)        & PC (S)         & PC (emp)       \\ \hline
\multicolumn{1}{|l|}{EV trip}  & 6.48 (1.49)    & 15.51 (1.49)  & 40.13 (0.80)   & 17.47          \\ \hline
\multicolumn{1}{|l|}{EV month} & 21.04 (2.17)   & 29.36 (4.59)  & 55.47 (1.64)   & 172.60         \\ \hline
\multicolumn{1}{|l|}{EV year}  & 143.88 (4.40)  & 184.22 (8.76) & 245.90 (5.00)  & 2131.36        \\ \hline
\multicolumn{1}{|l|}{HW trip}  & 11.88 (2.07)   & 18.55 (2.18)  & 48.59 (1.00)   & 24.47          \\ \hline
\multicolumn{1}{|l|}{HW month} & 81.79 (2.11)   & 127.69 (8.28) & 193.87 (12.34) & 293.70         \\ \hline
\multicolumn{1}{|l|}{HW year}  & 1052.06 (13.79) & 1576.40 (97.22) & 2226.76 (88.04) & 3565.91        \\ \hline
\end{tabular}
\vspace{2mm}
\caption{Average performance (seconds) and standard deviation over 10 runs. For MnC, the prover was a Raspberry Pi 4, Phone, or PC under Fast (F), Medium (M), or Slow (S) networks. For emp-zk, the prover was a PC with a Fast network.}

\label{tab:bench_results}
\end{table*}}

\section{Discussion}
\label{sec:discussion}
Recent advances in ZKP technology enable integration with identity systems~\cite{ernstberger2024you}. Our prototype using ZK-SecreC tooling shows that identifiers, hashes, and ECDSA signatures can be incorporated with vehicle, driver, and GPS records. Fraud and anomaly detection could also be addressed with limited disclosure techniques as in MPC~\cite{10.1007/978-3-662-47854-7_14}. While ZKPs provide strong privacy and integrity guarantees, integration must include secure logging without exposing sensitive data. With successful deployments in voting and cryptocurrency, we anticipate commercial-grade ZK-PoL systems within 3–5 years, contingent on investment.

Hardware requirements include a tamper-evident GPS module securely coupled to the vehicle, capable of signing trip data and distinguishing between normal driving and transport scenarios. Existing commercial devices partially meet these needs, typically offering sealed enclosures, secure key storage, or tamper alerts. While we assume the device cannot be undetectably removed or spoofed, this trust anchor remains a key deployment consideration, and additional safeguards—such as secure mounting or periodic verification—may be required in practice. Offloading ZKP computation to the module could improve usability but raises complexity. Our benchmarks (Section~\ref{sec:performance_evaluation}) indicate that annual proofs (43,800 points) require roughly 36 minutes (EV subsidy) or 59 minutes (highway taxation), though proofs can be batched. These runtimes are tolerable for infrequent proofs and can be amortized over idle or charging periods. Further reductions are expected from future work on non-interactive back-ends, rollups, and hardware acceleration.

ZK-PoL protocols meet emerging public sector needs for transparent subsidy distribution~\cite{iea} and help private entities prepare for regulatory shifts. A viable ecosystem will require coordination among service providers, vendors, and installers, plus interoperability standards to prevent fragmentation. By preserving user privacy and control~\cite{regulation2020art}, ZK-PoL offers a critical advantage as regulations evolve. Although proof generation demands user participation and computation time, these barriers could be mitigated by proof offloading, user-friendly apps, and back-end optimizations. Practical deployments will also need to address policy robustness, trust anchor management, and graceful failure handling.

\section{Conclusion and Future Work} 
\label{sec:conclusion}
This paper presented the prototyping of privacy-preserving mechanisms for vehicle subsidy and taxation compliance, making three main contributions. First, we formalized subsidy and taxation proof tasks in a way that remains close to societal intuitions while being compatible with ZKPs. Second, we demonstrated that existing ZK techniques can efficiently address these tasks. Finally, we analysed the steps and barriers toward practical deployment.

We introduced and evaluated ZK-PoL protocols that verify compliance claims while preserving detailed location privacy. Beyond mobility, similar protocols could secure high-value asset tracking in supply chains~\cite{nasrulin2018robust}, monitor smart city infrastructure~\cite{bornholdt2019proof}, verify attendance at events~\cite{pournaras2020proof}, and authenticate autonomous systems' locations in industrial settings~\cite{wu2024proof}. Applications in insurance are also promising, enabling verifiable incentives for low-risk driving and supporting usage-based models~\cite{mckinsey-insurance}.

Large-scale adoption will require overcoming infrastructure, scalability, and interoperability challenges. Addressing these issues is critical for integrating ZK-PoL systems into the evolving landscape of smart mobility and privacy-aware digital services.

\bibliographystyle{plain}
\bibliography{bibliography}

\appendix

\section{Additional algorithms in the Highway Tax use-case}\label{app:hwtax}

We collect here the specifications of local computations by the Prover. The first of them, given in Alg.~\ref{alg:findtriangle} determines whether a point is inside a triangle. One possible way to check this is to compute the area of the given triangle, and also compute the areas of three triangles that are formed by the given point and two vertices of the given triangle. If the given point is inside the given triangle, then these three triangles exactly cover the given triangle. If the given point is outside, then they also cover the given triangle, but also cover some area outside the triangle.

\begin{algorithm}
\caption{Finding the triangle: $\mathsf{find\_triangle}$}\label{alg:findtriangle}
\KwIn{Coordinates of a point $(\prepro{x},\prepro{y})$}
\KwIn{Vertices $(\prever{X_{j,k}}, \prever{Y_{j,k}})$ of triangles ($1\leq j\leq \prepub{n_\mathrm{tri}}$, $k \in \{1, 2, 3\}$)}
\KwOut{Index $\prepro{t}$ of the first triangle containing $(\prepro{x},\prepro{y})$, or $1$ if none}
\For{$i\leftarrow 1$ \KwTo $\prepub{n_\mathrm{tri}}$}{
    $\prever{a} \leftarrow\Delta\mathsf{area\_dbl}(\prever{X_{i,1}},\prever{Y_{i,1}},\prever{X_{i,2}},\prever{Y_{i,2}},\prever{X_{i,3}},\prever{Y_{i,3}})$\\
    $\prepro{b} \leftarrow\Delta\mathsf{area\_dbl}(\prever{X_{i,1}},\prever{Y_{i,1}},\prever{X_{i,2}},\prever{Y_{i,2}},\prepro{x},\prepro{y})$\\
    $\prepro{c} \leftarrow\Delta\mathsf{area\_dbl}(\prever{X_{i,1}},\prever{Y_{i,1}},\prepro{x},\prepro{y},\prever{X_{i,3}},\prever{Y_{i,3}})$\\
    $\prepro{d} \leftarrow\Delta\mathsf{area\_dbl}(\prepro{x},\prepro{y},\prever{X_{i,2}},\prever{Y_{i,2}},\prever{X_{i,3}},\prever{Y_{i,3}})$\\
    \If{$\prever{a} = \prepro{b} + \prepro{c} + \prepro{d}$}{
        $\prepro{t}\leftarrow i$\\
        \Return{$\prepro{t}$}
    }
}
\Return{1}
\end{algorithm}

The second local computation is that of $\mathsf{get\_bcoords}$, converting Cartesian coordinates to unnormalized barycentric coordinates (summing up to twice of the area of the triangle). The computation in Alg.~\ref{alg:findbarycoords} makes use of published formulas\footnote{\url{https://en.wikipedia.org/wiki/Barycentric\_coordinate\_system\#Vertex\_approach}} for conversion. In order to make sure that we get positive coordinates for points inside the triangle, we need to know the sign of the determinant in (\ref{eq:dblarea}). We thus define
\[
\Delta\mathsf{area\_dbl\_sgn}(a_1,b_1,a_2,b_2,a_3,b_3)=\mathrm{det}\left(\begin{matrix}
        a_1 & a_2 & a_3 \\
        b_1 & b_2 & b_3 \\
        1 & 1 & 1
\end{matrix} \right)\enspace.
\]
\begin{algorithm}
\caption{Barycentric coordinates of a point in a triangle: $\mathsf{get\_bcoords}$}\label{alg:findbarycoords}
\KwIn{Coordinates of a point $(\prepro{x},\prepro{y})$}
\KwIn{Coordinates of the vertices of a triangle: $(\prepro{a_1},\prepro{b_1}),(\prepro{a_2},\prepro{b_2}),(\prepro{a_3},\prepro{b_3})$}
\KwOut{Unnormalized barycentric coordinates $(\prepro{s},\prepro{t})$ of the given point with respect to the given triangle}
$\prepro{\mathit{area}}\leftarrow\Delta\mathsf{area\_dbl\_sgn}(\prepro{a_1},\prepro{b_1},\prepro{a_2},\prepro{b_2},\prepro{a_3},\prepro{b_3})$\\
$\prepro{sgn}\leftarrow\text{\leIf{$\prepro{\mathit{area}}\geq 0$}{$1$}{$-1$}}$\\
$\prepro{s}\leftarrow \prepro{sgn}*( \prepro{b_1}\cdot\prepro{a_3}-\prepro{a_1}\cdot\prepro{b_3}+(\prepro{b_3}-\prepro{b_1})\cdot\prepro{x}+(\prepro{a_1}-\prepro{a_3})\cdot\prepro{y}  )$\\
$\prepro{t}\leftarrow \prepro{sgn}*( \prepro{a_1}\cdot\prepro{b_2}-\prepro{b_1}\cdot\prepro{a_2}+(\prepro{b_1}-\prepro{b_2})\cdot\prepro{x}+(\prepro{a_2}-\prepro{a_1})\cdot\prepro{y}  )$\\
\Return{$(\prepro{s},\prepro{t})$}
\end{algorithm}

\section{Extended Security and Privacy Analysis} \label{app:security} 

We can present the security proof of our system in the universal composability (UC)~\cite{uc} framework (actually, our treatment is closer to the equivalent \emph{reactive simulatability} framework~\cite{brsim}). We can abstract the system model presented in Sec.~\ref{sec:system_model} as an ideal functionality, with clear interfaces for both the environment (a.k.a. ``the rest of the system'') and for the adversary. We can then show that the real system, consisting of the Witness device, the Prover, and the Verifier, all running the cryptographic technologies discussed in Sec.~\ref{sec:zk-specification}, is a secure implementation of this ideal functionality.

\subsection{Interface for the environment}\label{ssec:envinterface}

Our system conceptually consists of three components --- the Witness device, the Prover, and the Verifier ---, which react to the stimuli from the environment. Both the real system and the ideal functionality offer the same interface towards the environment. Logically, the interface is split into three parts, corresponding to the components. The environment can give the following ``stimuli'' (i.e. commands) to the real system or the ideal functionality, and receive the following replies:
\begin{itemize}
    \item At the beginning of the execution, the Witness device may receive the command $(\mathsf{init},\mathit{sid})$ from the environment. Here $\mathit{sid}$ is the ``session identifier'' that is used to tie together any other commands / messages that the system or functionality may exchange in response to this command.
    \item During the execution, the Witness device may receive commands of the form $(\mathsf{move},\mathit{sid},(x,y))$, where $(x,y)$ are the current coordinates into which the environment has presumably moved the Witness device.
    \item At some point during the execution, the Prover may receive $(\mathsf{prove},\mathit{sid},\vedom{\mathit{AD}})$, while the Verifier also receives $(\mathsf{verify},\mathit{sid},\vedom{\mathit{AD}})$. These commands indicate that the system is now supposed to check whether the coordinate trace satisfies some policy, a part of which is given by the \emph{authority's data} $\vedom{\mathit{AD}}$. The input $\vedom{\mathit{AD}}$ (which must be the same for the Prover and the Verifier) corresponds to Verifier's inputs to Alg.~\ref{alg:evsubsidy} and Alg.~\ref{alg:hwtax}. Note the use of $\mathit{sid}$ to indicate which request to prove and request to verify correspond to each other. In response to these commands, the environment receives $(\mathsf{ok},\mathit{sid})$ or $(\mathsf{not\_ok},\mathit{sid})$ from both the Prover and the Verifier.
\end{itemize}

\subsection{Ideal functionality}
Our ideal functionality $\mathcal{I}$ is parametrized with a relation $R$ corresponding to the \emph{policy} that the trajectory of the system has to satisfy. The arguments of $R$ are the list of coordinates $\prdom{\vec S}$ and the authority's data $\vedom{\mathit{AD}}$. Our ideal functionality is also parametrized with the hash function $H$ used in Alg.~\ref{alg:evsubsidy} and Alg.~\ref{alg:hwtax}.

The ideal functionality offers the described interface to the environment and another one to the adversary. It keeps the list of coordinates $\prdom{\vec S}$ as its internal state. The ideal functionality works as follows:
\begin{itemize}
    \item At the beginning of the execution, it may receive $(\mathsf{corrupt\_prover})$ or $(\mathsf{corrupt\_verifier})$ (but not both) from the adversary. If it receives them, $\mathcal{I}$ records that either Prover or Verifier has been corrupted.
    \item On input $(\mathsf{init},\mathit{sid})$ from the environment over Witness device's interface, $\mathcal{I}$ initializes the sequence of coordinates $\prdom{\vec S}$ to the empty sequence.
    \item On input $(\mathsf{move},\mathit{sid},\prdom{(x,y)})$ from the environment over Witness device's interface, $\mathcal{I}$ appends the coordinates $\prdom{(x,y)}$ to the sequence of coordinates $\prdom{\vec S}$.
\end{itemize}
These steps are straightforward. The handling of the proof request is more complicated, because the output depends not only on the policy $R$ and the list of coordinates $\prdom{\vec S}$, but also on authority's data $\vedom{\mathit{AD}}$, whether the same $\vedom{\mathit{AD}}$ was input to the Prover and the Verifier, and, in case the Prover or the Verifier is corrupted, also on the choices of the adversary. Thus, 
    on input $(\mathsf{prove},\mathit{sid},\vedom{\mathit{AD}_P})$ from the environment to the Prover, and $(\mathsf{verify},\mathit{sid},\vedom{\mathit{AD}_V})$ to the Verifier, $\mathcal{I}$ works as follows:
    \begin{itemize}
        \item Initialize $\mathit{res}_P\leftarrow 1$ and $\mathit{res}_V\leftarrow 1$
        \item Send $H(\prdom{\vec S})$, $\vedom{\mathit{AD}_P}$, and $\vedom{\mathit{AD}_V}$ to the adversary.
        \item If Prover is corrupted, then also send $\prdom{\vec S}$ to the adversary.
        \item If $R(\prdom{\vec S},\vedom{\mathit{AD}_P})=0$, then put $\mathit{res}_P := 0$.
        \item If $\vedom{\mathit{AD}_P}\not=\vedom{\mathit{AD}_V}$, or if $R(\prdom{\vec S},\vedom{\mathit{AD}_V})=0$, then put $\mathit{res}_V := 0$.
        \item If Prover is corrupted, then send $(\mathsf{output},\mathit{sid},\mathit{res}_P)$ to the adversary, and receive back $(\mathsf{update},\mathit{sid},b)$. Update $\mathit{res}_P := b$.
        \item If Verifier is corrupted, then send $(\mathsf{output},\mathit{sid},\mathit{res}_V)$ to the adversary, and receive back $(\mathsf{update},\mathit{sid},b)$. Update $\mathit{res}_V := b$.
        \item If $\mathit{res}_P=0$, then send $(\mathsf{not\_ok},\mathit{sid})$ to the environment over Prover's connection to the environment, otherwise send $(\mathsf{ok},\mathit{sid})$ over Prover's connection.
        \item If Prover is corrupted, then wait for $(\mathsf{proceed},\mathit{sid})$ from the adversary.
        \item If $\mathit{res}_V=1$, then send $(\mathsf{ok},\mathit{sid})$ to the environment over Verifier's connection to the environment, otherwise send $(\mathsf{not\_ok},\mathit{sid})$ over Prover's connection.
    \end{itemize}
We see that there are some values that $\mathcal{I}$ does not attempt to keep private. We have discussed previously that we consider the leak of $H(\prdom{\vec S})$ acceptable. We also see that the adversary determines, what a corrupt Prover or Verifier returns to the environment; this design choice is a standard one. Finally, we see that if $\prdom{\vec S}$ does not satisfy the policy, then there is no way for an uncorrupted Verifier to return $\mathsf{ok}$.

\subsection{Real system}

The real system consists of (the Turing machines realizing the steps of) the Witness device, the Prover and the Verifier. All machines are parametrized with the hash function $H$. The Prover and the Verifier are also parametrized with the relation $R$.

The Prover and the Verifier implement the ZKP protocol for the relation $R$. In order to compartmentalize it, we make use of the composability properties of the UC framework. Namely, we introduce a fourth machine $\mathcal{F}_\mathrm{ZK}^R$ to the real system. This machine is an ideal functionality modelling a zero-knowledge protocol for the relation $R$. The functionality is defined in~\cite[Fig.~11]{zksc-CSF24}, specialized from the functionality given in~\cite{canetti-fischlin-uccom}. It accepts inputs $(\mathsf{prove!},\mathit{sid},x,w)$ from the Prover machine and $(\mathsf{prove?},\mathit{sid},x')$ from the verifier machine, and sends back $(\mathsf{proven},\mathit{sid})$ to the Verifier machine if $R(x,w)$ holds, $x=x'$, and the adversary allows it to proceed. Given the description of $R$ as an arithmetic circuit, there are a number of protocols that provide a secure implementation of $\mathcal{F}_\mathrm{ZK}^R$. These secure implementations consist of a prover machine for ZKP, and a verifier machine for ZKP; in the actual deployment, these machines would be a part of our Prover and Verifier.

\subsubsection{Witness device}
The Witness device maintains the list of coordinates $\prdom{\vec S}$. It works as follows:
\begin{itemize}
    \item On input $(\mathsf{init},\mathit{sid})$ from the environment, generates a key pair $(\mathit{pk},\mathit{sk})$ for signing. Sends $(\mathsf{witpk},\mathit{sid},\mathit{pk})$ to the Prover and the Verifier. Initializes the sequence of coordinates $\vec S$ to the empty sequence.
    \item On input $(\mathsf{move},\mathit{sid},\prdom{(x,y)})$ from the environment, appends the coordinates $\prdom{(x,y)}$ to the sequence of coordinates $\prdom{\vec S}$.
    \item On input $(\mathsf{getcoords},\mathit{sid})$ from the Prover machine, sends $(\mathsf{coords},\mathit{sid},\prdom{\vec S},\sigma)$ back to the Prover machine, where $\sigma=\mathsf{sig}_\mathit{sk}(H(\prdom{\vec S}))$.
\end{itemize}
We see that the description of the Witness device very much matches our system model in Sec.~\ref{sec:system_model}.

\subsubsection{Prover machine} 
The Prover machine gets the list of coordinates from the Witness device, and uses them to present a ZKP of policy satisfaction to the Verifier (making use of $\mathcal{F}_\mathrm{ZK}^R$). It works as follows:
\begin{itemize}
    \item On input $(\mathsf{witpk},\mathit{sid},\mathit{pk})$ from the Witness device, store $\mathit{pk}$.
    \item On input $(\mathsf{prove},\mathit{sid},\vedom{\mathit{AD}_P})$ from the environment, send $(\mathsf{getcoords},\mathit{sid})$ to the witness device, and expect back $(\mathsf{coords},\mathit{sid},\prdom{\vec S},\sigma)$. Verify the signature $\sigma$ on $H(\prdom{\vec S})$ using the public key $\mathit{pk}$, and check whether $R(\prdom{\vec S}, \vedom{\mathit{AD}_P})$ holds. If not, send $(\mathsf{not\_ok},\mathit{sid})$ to the environment. If yes, send $(\mathsf{sig},\mathit{sid},H(\vec S),\sigma)$ to the verifier machine, and $(\mathsf{prove!},\mathit{sid},(\vedom{\mathit{AD}_P}, H(\vec S)),\vec S)$ to $\mathcal{F}_\mathrm{ZK}^R$.
\end{itemize}
The adversary may corrupt the Prover machine by sending it the $(\mathsf{corrupt})$-command at the beginning of the execution. In this case, the Prover machine sends $\vedom{\mathit{AD}_P}$, $\prdom{\vec S}$ and $\sigma$ also to the adversary. Also, the adversary controls, what the Prover machine sends to the Verifier, to $\mathcal{F}_\mathrm{ZK}^R$, and to the environment.

\subsubsection{Verifier machine}
The Verifier machine accepts the proof from the Prover machine (via $\mathcal{F}_\mathrm{ZK}^R$) It works as follows:
\begin{itemize}
    \item On input $(\mathsf{witpk},\mathit{sid},\mathit{pk})$ from the witness device, store $\mathit{pk}$.
    \item On input $(\mathsf{verify},\mathit{sid},\vedom{\mathit{AD}_V})$ from the environment, expect $(\mathsf{sig},\mathit{sid},h,\sigma)$ also from the Prover machine. Verify $\sigma$ on $h$, using the public key $\mathit{pk}$. If it did not verify, send $(\mathsf{not\_ok},\mathit{sid})$ to the environment. If it verified, then send $(\mathsf{prove?},\mathit{sid},(\vedom{\mathit{AD}_V},h))$ to $\mathcal{F}_\mathrm{ZK}^R$. On input $(\mathsf{proven},\mathit{sid})$ from $\mathcal{F}_\mathrm{ZK}^R$, send $(\mathsf{ok},\mathit{sid})$ to the environment.
\end{itemize}
Verifier machine may be corrupted. In this case, the adversary controls what the verifier machine sends to $\mathcal{F}_\mathrm{ZK}^R$ and to the environment. Also in this case, the Verifier machine sends $\vedom{\mathit{AD}_V}$ and $h$ to the adversary.


\subsection{Simulators}

The goal of this section is to show that the real system is \emph{at least as secure as} the ideal functionality. We have to show that for any environment $\mathcal{Z}$ connecting to the interface described in Sec.~\ref{ssec:envinterface} of the real system, and for any adversary $\mathcal{A}$ connecting to the adversarial interface of both the real system and the environment $\mathcal{Z}$, there exists an adversary $\mathcal{S}$ connecting to the adversarial interface of both the ideal functionality $\mathcal{I}$ and the environment $\mathcal{Z}$, such that the environment $\mathcal{Z}$ cannot distinguish whether it is executing together with the real system and $\mathcal{A}$, or with the ideal functionality and $\mathcal{S}$.

Given $\mathcal{A}$, we construct $\mathcal{S}$ as the composition $\mathit{Sim}\|\mathcal{A}$, where $\mathit{Sim}$ is an interactive Turing machine. In the rest of this section, we give the description of $\mathit{Sim}$ and argue that it indeed makes $\mathit{Sim}\|\mathcal{A}$ a suitable adversary $\mathcal{S}$. Note that the machine $\mathit{Sim}$ basically has two interfaces:
\begin{itemize}
    \item At one side, it is able to connect to the interface that $\mathcal{I}$ offers to the adversary.
    \item At the other side, it must provide the same interface to $\mathcal{A}$ that is provided by the real system.
\end{itemize}
In order to show that the simulator $\mathit{Sim}$ turns any $\mathcal{A}$ to a suitable $\mathcal{S}$, we have to show that the real system is indistinguishable to the composition of the ideal functionality and the simulator. This indistinguishability must be for a distinguisher that connects to both the interface for the environment (Sec.~\ref{ssec:envinterface}), as well as to the interface for $\mathcal{A}$ provided by the real system.

The simulator ``simulates'' the executions of uncorrupted parties, perhaps with less inputs than the actual parties would have. If neither the Prover nor the Verifier have been corrupted, then the construction of the simulator is trivial. In this case, the simulator receives $H(\prdom{\vec S})$, $\vedom{\mathit{AD}_P}$ and $\vedom{\mathit{AD}_V}$ from $\mathcal{I}$. It does not have to send any messages itself, except for notifying $\mathcal{A}$ that a ZK proof is currently running by $\mathcal{F}_\mathrm{ZK}^R$.

If either the Prover or the Verifier has been corrupted, then the simulator has to construct the messages that the corrupted party would send to the adversary $\mathcal{A}$. Also, the simulator has to be able to handle the messages that the corrupted party sends out; these messages have been constructed by $\mathcal{A}$. When $\mathcal{A}$ sends out the corruption request, the simulator forwards it to $\mathcal{I}$. The following simulation depends on which party was corrupted. As the request comes out at the beginning of the execution, we can split the construction of the simulator to two parts.

\subsubsection{Simulator for corrupted Verifier}

During the initialization, the simulator generates a public-private key pair $(\mathit{pk},\mathit{sk})$. Sends $(\mathsf{witpk},\mathit{sid},\mathit{pk})$ to the adversary (as Witness device sending a message to the Verifier).

During proving, simulator gets $\vedom{\mathit{AD}_P}$, $\vedom{\mathit{AD}_V}$ and the hash $h$ (and $\mathit{sid})$ from the ideal functionality. Signs $h$ using $\mathit{sk}$, obtaining $\sigma$. Also gets the value $\mathit{res}_V$ from the ideal functionality. Sends $(\mathsf{sig},\mathit{sid},h,\sigma)$ to the adversary (as Prover sending a message to the Verifier). Gets back $(\mathsf{prove?},\mathit{sid},(\vedom{\mathit{AD}_V},h))$ as a message from the Verifier to $\mathcal{F}_\mathrm{ZK}^R$. If the the value $h$ differs from its previous value, or if $\vedom{\mathit{AD}_V}$ differs from $\vedom{\mathit{AD}_P}$ received previously, then $\mathit{Sim}$ sets $\mathit{res}_V:=0$. Sends $(\mathsf{update},\mathit{sid},\mathit{res}_V)$ to the ideal functionality.

We see that if the corrupted Verifier (controlled by the adversary) follows the protocol, submitting $h$ and $\vedom{\mathit{AD}_V}$ to the ZK proof system, then the output that the environment receives from the Verifier is the same with both the real system and with the ideal functionality, depending on the values $\prdom{\vec S}$ and $\vedom{\mathit{AD}_P}$ that the environment submits through Prover's interface. If the corrupt Verifier changes the values it gives to $\mathcal{F}_\mathrm{ZK}^R$, then the output to the environment will be $\mathsf{not\_ok}$.

\subsubsection{Simulator for corrupted Prover}

During the initialization, the simulator generates a public-private key pair $(\mathit{pk},\mathit{sk})$. Sends $(\mathsf{witpk},\mathit{sid},\mathit{pk})$ to the adversary (as Witness device sending a message to the Prover).

During proving, $\mathit{Sim}$ gets $\prdom{\vec S}$, $\vedom{\mathit{AD}_P}$, and $\vedom{\mathit{AD}_V}$ from the ideal functionality $\mathcal{I}$. Computes $\sigma$ on $\prdom{\vec S}$, using $\mathit{sk}$. Receives $(\mathsf{getcoords},\mathit{sid})$ from the adversary (as Prover sending a message to the Witness device). Sends $(\mathsf{coords},\mathit{sid},\prdom{\vec S},\sigma)$ back. Will again get these values from the adversary (as Prover submitting its inputs to $\mathcal{F}_\mathrm{ZK}^R$). Will also get the value $b$ that the adversary $\mathcal{A}$ wants the Prover machine to return to the environment. Sends it to the ideal functionality as $(\mathsf{update},\mathit{sid},b)$. If the adversary has sent a different $\vedom{\mathit{AD}_P}$ to $\mathcal{F}_\mathrm{ZK}^R$, or it has changed $\prdom{\vec S}$, then do not send $(\mathsf{proceed},\mathit{sid})$ to the ideal functionality.

The difference of the real system, and the composition of the ideal functionality and simulator is the following. In the real system, $\mathcal{F}_\mathrm{ZK}^R$ is used to evaluate $\prdom{\vec S}$ and \textit{authority's data}. This evaluation is done by executing $R$ on these data, and the hash function and the signature are used to protect the integrity of the witness that goes into $\mathcal{F}_\mathrm{ZK}^R$. In the ideal system, the ideal functionality will evaluate the policy on $\prdom{\vec S}$ and $\vedom{\mathit{AD}}$; the integrity of $\prdom{\vec S}$ holds by default because it is not moved around before policy evaluation. Hence we have to argue that $R$ is a correct implementation of the policy. We also have to argue that the probability of using different $\prdom{\vec S}$ in the real system and in the composition of the ideal system and the simulator simulator, is negligible.

We have argued that $R$ is a correct implementation of the policy in Sec.~\ref{subsec:ev-protocol-description} and in Sec.~\ref{subsec:hwtax-protocol-description}. Also, changing $\prdom{\vec S}$ in the real system requires the adversary to either forge the signature or find a different pre-image of the hash function.

\end{document}